 \documentstyle[aps,pre]{revtex}          
\tighten                                  
\begin{document}                          
 \draft                                   
 \twocolumn                               

\title{Flame propagation in random media}

\author{N. Provatas,$^{1}$ T. Ala-Nissila,$^{1,2,3}$
 Martin Grant,$^{1}$ K. R. Elder,$^{1}$ and Luc Pich\'e$^{1,4}$}

\address{
$^1$McGill University
Physics Department and Centre for the Physics of Materials,
3600 rue University, Montr\'eal, Qu\'ebec, Canada H3A 2T8
}

\address{
$^2$University of Helsinki, Research Institute for Theoretical Physics,
\\P.O. Box 9 (Siltavuorenpenger 20 C), FIN--00014 university of Helsinki,
Finland
}

\address{
$^3$Brown University, Department of Physics, Box 1843,
Providence, R.I. 02912, U.S.A.
}

\address{
$^4$National Research Council, Industrial Materials Institute,
75 De Mortagne Boulevard, \\
Boucherville, Qu\'ebec, Canada, J4B 6Y4
}

\date{\today}

\maketitle
\narrowtext

\begin{abstract}
We introduce a phase-field model to describe the dynamics
of a self-sustaining propagating combustion front within a medium of
randomly distributed reactants. Numerical simulations of this
model show that a flame front exists for
reactant concentration $c > c^* > 0$, while
its vanishing at $c^*$ is consistent with mean-field percolation
theory.  For $c > c^*$, we find that the interface associated
with the diffuse combustion zone exhibits kinetic roughening
characteristic of the Kardar-Parisi-Zhang equation.
\end{abstract}



The behavior of a nonequilibrium system is often limited by a reaction
front between one metastable or unstable phase and another more stable
phase.  A common but spectacular example of this occurs in slow
combustion where a flame front forms and can propagate
\cite{Wil85,Alb86,Bak90,Dro92,Bak87,Zha92}.  Here we study a
phase field model of combustion in the absence of convection and
apply methods developed in the study of phase transitions
to the asymptotic behavior of a self-sustaining combustion
front growing within a medium of randomly distributed reactants.
Both the formation of the front as well as its universal dependence
on length and time are examined.  We find numerically that
the combustion front exists for reactant
concentration $c > c^* >0 $ in $d=2$ dimensions, where the behavior at
$c^*$ is consistent with that of mean-field percolation.  For $c >
c^*$, we find that the diffuse combustion zone exhibits kinetic
roughening characteristic of the Kardar-Parisi-Zhang interface equation
\cite{KPZ}.

While we expect our phase-field model to describe the universal features of
combustion or reaction fronts in the absence of convection we shall
motivate it below in terms of a specific example, forest fires.
The physics associated with forest fires has recently received
attention \cite{Alb86,Bak90,Dro92} due to the potential relationship
with the concept of self-organized criticality,
introduced by Bak \cite{Bak87} and collaborators.  In most cases
studied to date, cellular automaton models on a lattice have been
used \cite{Alb86,Bak90,Dro92}.   In these works a
collection of forests which can burn and subsequently reappear is
considered.  In contrast, this paper focuses on systems
in which the reacting element cannot spontaneously reappear.

Our model consists of two coupled reaction-diffusion
equations, one governing the concentration of reactants, and the
other the dynamics of a thermal field.  Unlike discrete lattice models
this model incorporates the interplay between long-range thermal
diffusion and local random concentration fields.  Within our model,
variations in the local temperature field $T(\vec x,
t)$ at position $\vec x$ and time $t$, are due to three effects:
(i) thermal diffusion through the medium; (ii) Newtonian cooling due to
coupling to a heat bath; and (iii) generation of heat, limited by
activation, from the reactants.  Explicitly, the temperature field obeys
\begin{equation}
\frac{\partial T}{\partial t} = {D} \nabla^2 T
-{\Gamma}[T-T_0]-  \vec V  \cdot {\nabla} T + R(T,c),
\label{add4}
\end{equation}
where $D$ is the diffusion coefficient, $\Gamma$ is the thermal
dissipation constant, and $T_0$ is the constant background temperature
of the bath responsible for Newtonian cooling.  For completeness we
have included convection due to an external forcing $\vec V$, but we
shall hereafter set this term to zero.

Nonlinearities enter through the reaction rate $R(T,c)$, which is
limited by the local concentration of reactants $c$.   Activation
implies that this term is proportional to $\exp (-A/T)$, where $A$ is
an activation energy, and Boltzmann's constant has been set to unity.
Furthermore the rate is limited by the local flux $\propto \sqrt{T}$,
so that the local energy produced per unit time is $q(T) = T^{3 / 2}
\exp(- A/ T)$, where the additional factor of ${T}$ sets the scale of
energy.  In the combustion literature \cite{Wil85},
the multiplicative prefactor is
$T^\alpha$, where $\alpha={ {\cal O} (1)}$ varies according to the
conditions present.   However,
the particular form is relatively unimportant
since the dominating  temperature dependence is the exponential
activation.  Hence, on measuring temperature in units of the activation
energy, we model the reaction rate as
\begin{equation}
 R = \lambda_2 \biggl{(} T^{3/2} e^{-1/T}\biggr{)}  c
= -\lambda_1 \frac{\partial c}{\partial t},
\label{2}
\end{equation}
where $\lambda_1$ is a dimensionless constant.  This  completes
the formulation.  Initial conditions determine the random
distribution and concentration of reactant.  Here we have
initially distributed the reactant at random, with the
probability that a given system site is occupied being $c$.

For the remainder of this paper, we consider $d = 2$, where a front
initially parallel to the $y$ axis propagates in the $x$ direction.
The dimensionless parameters are set to $D=0.2$, $\Gamma=0.05$, $T_0=0.01$ and
$\lambda_1=8$, and time is measured in units of those for the reaction,
$\lambda_2/\lambda_1$, and length in units of the dimension of the reactant.
In our numerical work, the mesh size in space is set to
$\Delta x = 1$, while the mesh size in time is $\Delta t = 0.01$;
tests of smaller mesh sizes give qualitatively similar results.
It is useful to relate these choices of parameters to the specific
example of a forest fire.  For example, the constant
$\lambda_2$ in Eq.\ 2, can be found in terms of the density and
specific heat of air and the activation temperature
of wood \cite{parameters}.  In physical units, we have
$D\sim 1 m^2 s^{-1}$, $\Gamma \sim 0.05 s^{-1}$, $T_0\sim 10K$
and $c_p\sim  5J g^{-1} K^{-1}$ and $A \sim 500 K$.  With the exception
of $T_0$, these are comparable to real systems.  Our small $T_0$ has
been chosen to give enhanced cooling and hence keep diffusion fields
relatively short ranged.  This allows us to perform our numerical
integrations with good accuracy without having to simulate extremely
large systems.  Test runs show that our results are relatively
insensitive to the choice of $T_o$.

Due to the activated nature of the combustion process, a
self-sustaining propagating combustion front requires a sufficient
amount of heat to be released during combustion.  The source for this
heat is the reactant concentration $c$.  Since activation limits
the production of this heat, we expect existence of a nonzero
concentration $c^*$, below which the fire will spontaneously burn out
due to insufficient heat production.  That is, for $c < c^*$ the
velocity of the front $v(c) = 0$, while $v$ is nonzero for higher
concentrations.  For quantitative analysis the position $h(x,t)$
of the front, where $v = \partial h/\partial t$, is defined as the
position $x$ where the temperature is maximum at a given time and
$y$ position.  The variable $h$ is then a single-valued function of
$y$.

For large concentrations, $v$ is constant after an initial
transient, and decreases with $c$.  The transient increases
as $c^*$ is approached.  In the vicinity of $c^*$,
the asymptotic velocity approaches the relationship
$v(c) \sim (c-c^*)^\phi$, where $\phi$ is an exponent similar
to that obtained in percolation theory \cite{Sta79}.

To determine the scaling exponent in the case of a random background,
Eqs.\ (2) and (3) were numerically solved on a lattice using
periodic boundary conditions in the $y$ direction and fixed boundary
conditions in the $x$ direction.  The dimension of the system is $L$ in
the $y$ direction, and well exceeds $vt_{\rm max}$ in the $x$
direction, where $t_{\rm max}$ is the maximum time studied.  At every
site, the density variable $c$ is initially either zero or one, and
randomly
distributed with average $c$. The fire is started at the far left by
igniting a complete row of ``trees'' at $y=0$. After a short transient,
the propagating fire front assumes a steady-state average velocity
$v(c)$.  In Figs.\ 1(a) and (b) typical configurations of the
propagating temperature field are shown at $c=0.65$ and 0.225.
For lower densities, the front becomes very irregular and finally stops
propagating.  Calculating the velocity numerically for $L=200$ we find
$c^*=0.19 \pm 0.02$ and $\phi=0.46 \pm 0.09$.

Mean-field theory is useful to understand these
percolation results.
Consider a uniform distribution of reactants, whose density
variable $c$ is every where equal to a constant.
In this description there are no longer
variations in $T$ in the $y$ direction.
Assume there exists a mean-field temperature front $T_m$ moving with
constant velocity $v_m(c)$.
Using $\partial T/\partial t = v_m dT/dx$, we
obtain a nonlinear consistency
relation between $T_m$ and $v_m(c)$.
We have solved this mean field model numerically, revealing a
dependence of $v_m(c)$ in $c$ of the form $v_m(c) \propto
(c-c^*)^\phi$, near $c^*=0.19$, where $\phi=0.5$.
The exponent $\phi=1/2$ is also a consequence of the
following argument.
Let $x_t$ be a point sufficiently ahead of the peak of $T_m$,
beyond which dissipation exceeds activation in the thermal
diffusion equation.
Expanding $R(T)$ to linear order around $T_{tail} \equiv T_m(x_t)$, we
find that the leading edge of $T_m$ goes as
$T_m \sim \exp[(-v_m/2D - \sqrt{v_m^2 -
4D(\lambda_1 c q^{\prime}(T_{tail}) - \Gamma}) x]$.
The requirement that $T_m$ does not develop any oscillatory
components gives
$ v_m \ge 4 D \lambda_1 q^{\prime} (T_{tail}) ( c - c^* )^{1/2} $,
with $ c^* = \Gamma / ( q^{\prime} (T_{tail}) \lambda_1)$.
This analysis is analogous to the method used in Ref. \cite{Mur89} to
find front velocities in the context of epidemic models.

To incorporate finite-size effects, we use a scaling form
\begin{equation}
v(c,L) \sim L^{-\phi/ \nu } \Omega [(c-c^*) L^{1/ \nu} ].
\label{5}
\end{equation}
This is the same as that used in percolation theory \cite{Sta85}.  Here $\nu$
is the correlation-length exponent $\xi \sim (c-c^*)^{-\nu}$, and the
scaling function $\Omega(x \rightarrow \infty) \sim x^\phi$.  In
Fig.\ 2, we show numerical results for $v(c,L) L^{\phi /\nu}$ .vs.\
$(c-c^*) L^{1/ \nu}$ for nine different system sizes ranging from $L=4$
to $L=200$.  Using $c^*=0.19$ and $\phi=0.46$, we find that the best
collapse occurs for $\nu=0.6 \pm 0.1$, as shown in Fig.\ 2.

The results for the critical exponents are consistent
with the mean field exponents of percolation, for which $\phi=\nu=1/2$
\cite{Sta79}.  Qualitatively, heat propagation in our model is limited
by a percolation lattice, provided by the random density field $c$.
Below $c^*$, the connected cluster available for front propagation
breaks down, and the fire spontaneously dies out. The mean-field nature
of the critical exponents is due to the long-range nature of the
diffusion field associated with $T$.

For $c > c^*$, it is clear from Fig.\ 1 that the propagating interface
associated with $T$ develops large fluctuations and appears rough.  We
define the width of the interface by $w=\langle (h-\langle h\rangle )^2
\rangle^{1/2}$.  Rough interfaces often satisfy the scaling relation
\cite{KPZ,Fam85} $w(t,L) \sim t^\beta f(t/L^z)$ for large $L$ and $t$.
An important example of this is the Kardar-Parisi-Zhang (KPZ) interface
equation \cite{KPZ}, for which the exact and nontrivial exponents are
$\beta=1/3$, $z=3/2$, and $\chi=z\beta=1/2$, in $d=2$.  In our case,
for any given value of $c > c^*$, we expect the width to obey this
scaling form, i.e., for large $t$, we expect $w \sim t^{\beta}$ in the
limit $t \ll L^z$, where finite-size effects can be
neglected.  To account for the dependence of the width on concentration
in this limit, we propose
\begin{equation}
w(c,t)=\xi(c) w_s(t/\tau(c)),
\end{equation}
where
$w_s(t \rightarrow \infty) \sim t^{\beta}$, and for $c$ near $c^*$, $\xi(c)
\sim (c-c^*)^{-\nu}$ and  $\tau(c) \sim (c-c^*)^{-\Delta}$, where
$\Delta$ is a slowing down exponent.  In mean field, $2\nu = \Delta =
1$.  It is straightforward to generalize this form to include
finite-size effects.  In Fig.\ 3 we show the scaled width $w_s$ plotted
.vs.\ the scaled time $t_s=t/\tau$,
for seven different values of $c$ with $L = 200$.  For
this $L$, finite-size effects play no discernible role.  The
inset shows the original data set. A transient time $t_0$ has
been subtracted, which has been determined from the point where $v(c)$
reaches a constant value.  From the fitted $\xi(c)$ and  $\tau(c)$
for the data collapse we cannot accurately estimate $\nu$ and $\Delta$,
although they are consistent with the mean-field values.

{}From
the scaled data of Fig.\ 3, we determine the roughening exponent
$\beta$. The running slope of the data from a $\log w$ vs.\  $\log t$
plot gives an effective $\beta(t)$, which is shown in Fig.\ 4.  After
an initial transient the slope clearly tends towards $\beta =  1/3$,
which is the exact KPZ value.  We have also analyzed the data by
calculating the difference $w(bt)-w(t)=A (b^\beta -1) t^\beta$, where
$b$ is a constant (e.g.,  $b=2$). From this we find $\beta=0.34 \pm
0.04$, which is our best estimate of this exponent.

In the limit where the width saturates due to finite-size effects,
i.e., $t \gg L^z$,
the scaling relation gives $w(c,L) \sim L^\chi$.  Using system sizes
$L=50$, 76, 100, 150, 200, 300, 400 and 600, we obtain $\chi=0.5\pm 0.1$
for $c=0.5$ and $\chi=0.5\pm 0.2$ for $c=0.85$, as compared to the
exact KPZ value of $\chi = 1/2$.  Our results for $\beta$ and $\chi$
are therefore in good agreement with those of the KPZ equation
\cite{KPZ,Zha92}.

To summarize, in this work we have developed a realistic phase-field
model for combustion fronts.  We find a percolation transition at a
critical density $c^*\approx 0.19$, below which the fire will
spontaneously die. We have analyzed the nature of this transition and
found critical exponents which are consistent with those of mean-field
percolation. Furthermore, above $c^*$, we found that the diffuse
combustion front displays kinetic roughening.  By an appropriate
generalization of the usual scaling form for interfaces, we have shown
that the roughening exponents are compatible with the KPZ universality
class.

The Centre for Scientific Computing (CSC) Co.\ of Espoo, Finland has
provided most of the computing resources for this work.  This work has
also been supported by the Academy of Finland, the Natural Sciences and
Engineering research Council of Canada and {\it les  Fonds pour la
Formation de Chercheurs et l'Aide \`a la Recherche de Qu\'ebec\/}.

\begin{figure}
\caption{
The temperature field $T(x,y,t)$
for a moving fire front in a uniform, random forest with
(a) $c=0.65$, and (b) $c=0.225$. The black pixels correspond to
the temperature field; the higher the temperature
the blacker the pixels. The interface $h(x,t)$ is
defined by the curve outlined by the
blackest pixels.  The light grey pixels to the right of the interface
represent reactant.
}
\end{figure}

\begin{figure}
\caption{
Finite size scaling of $v(c,L)$.  The main figure
shows $\rm ln ( v(c,L) l^{\phi / \nu} )$ .vs.\ $\rm ln( (c-c^*)L^{1 / \nu} )$.
The inset shows the unscaled data for system sizes
$L=4,6,8,24,44,54,64,104,200$, from right to left.
Sizes larger than $L=24$ lie almost on the same curve.
}
\end{figure}

\begin{figure}
\caption{
Crossover scaling function $w_s$ plotted .vs.\ $t_s=t / \tau$.
The inset shows the concentration dependent
width $w(c,t)$, for $c=0.4,0.5,0.6,0.7,0.75,0.80$
and $0.85$ from top to bottom.  The roughness increases with decreasing
density.  A transient time $t_o$ and the corresponding
offset $w_o$ has been subtracted from each $w(c,t)$
}
\end{figure}

\begin{figure}
\caption{
A log-log plot of the scaling function $w_s$ of Fig.\ 3.  The inset shows the
effective $\beta$ as a function of time.  The straight line represents
$\beta = 1/3$.
}
\end{figure}


\begin{thebibliography}{99}
\bibitem{parameters}
For a rough estimate of the parameters entering our model for
the specific case of a forest fire, we assume trees burn via a
steady-state reaction with air, treating air as an ideal gas.  The flux
of molecules striking the wood surface is $N_s=(n/4) \sqrt(2T/m)$,
where $n$ is the number density, and $m$ is the mass per molecule in
air.  Since the number of molecules reacting is $N_r=N_s e^{-A/T}$, the
energy $Q$ released per unit tree area and time equals the energy of
the molecules entering the reaction, $Q=(3T/2) N_s e^{A/T}$.  If $r$ is
a tree radius, the heat given off per unit volume of a tree site is $2
Q /r$.  Writing $T$ in units of $A$, the temperature change per unit
time corresponding to this heat production is $(2 Q/r)/(c_p \rho)$,
where $c_p$ is the specific heat and $\rho$ is the mass density of
air.  Now, on recalling that we are scaling temperature in units of
$A$, we obtain the parameter, $\lambda_2=(3 \sqrt{2} /4 r c_p)
(1/m)^{3/2} (A)^{1/2}$.
\end{thebibliography}
\end{document}